\documentclass[preprint,aps,prc,showpacs,tightenlines,preprintnumbers,amsmath,amssymb]{revtex4}

\usepackage[utf8]{inputenc}
\usepackage{graphicx}
\usepackage{bm}                      
\usepackage{mathptmx}                
\usepackage{newtxmath}
\usepackage{amssymb}  
\usepackage{dcolumn}                 
\usepackage{color}
\usepackage{comment}
\usepackage{natbib}

\newcommand{\be}{\beta}

\newcommand{\beq}{\begin{equation}}
\newcommand{\eeq}{\end{equation}}
\newcommand{\ba}{\begin{array}}
\newcommand{\ea}{\end{array}}
\newcommand{\bea}{\begin{eqnarray}}
\newcommand{\eea}{\end{eqnarray}}
\newcommand{\bi}{\begin{itemize}}  
\newcommand{\ei}{\end{itemize}}
\newcommand{\ben}{\begin{enumerate}} 
\newcommand{\een}{\end{enumerate}}
\newcommand{\bc}{\begin{center}}
\newcommand{\ec}{\end{center}}
\newcommand{\scro}{Scr\"oedinger }

\newcommand{\ee}[1]{\times 10^{#1}}

\def\bea{\begin{eqnarray}}
\def\eea{\end{eqnarray}}
\def\be{\begin{equation}}
\def\ee{\end{equation}}

\def\beq{\begin{equation}}
\def\eeq{\end{equation}}
\def\bar{\begin{array}[b]}
\def\barc{\begin{array}}
\def\bart{\begin{array}[t]}
\def\ear{\end{array}}

\begin{document}

\title{Structural aspects of FRG in quantum tunnelling computations}

\author{Alfio Bonanno}
\email{alfio.bonanno@inaf.it}
\affiliation{INAF, Osservatorio Astrofisico di Catania, via S.Sofia 78, 
I-9 5123 Catania, Italy;\\
INFN, Sezione di Catania, via S. Sofia 64, 
I-95123,Catania, Italy}

\author{Alessandro Codello}
\email{acodello@fing.edu.uy}
\affiliation{Instituto de F\'isica, Faculdad de Ingenier\'ia, Universidad de la Rep\'ublica, 11000 Montevideo, Uruguay}

\author{Dario Zappal\`a}
\email{dario.zappala@ct.infn.it}
\affiliation{
INFN, Sezione di Catania, Via Santa Sofia 64, 95123 Catania, Italy}

\begin{abstract}
\centerline{\bf ABSTRACT}
We probe  both the unidimensional quartic harmonic oscillator 
and the double well potential through a numerical analysis of the 
Functional Renormalization Group flow equations truncated at first order
in the derivative expansion. The two partial differential 
equations for the potential 
$V_k(\varphi)$ and the wave function renormalization $Z_k(\varphi)$,
as obtained in different schemes and with distinct regulators, 
are studied down to $k=0$, and the energy gap between lowest and first 
excited state is computed, in order to test the reliability of 
the approach in a strongly non-perturbative regime. Our findings 
point out at least three ranges of the quartic coupling $\lambda$, 
one with higher $\lambda$ where the 
lowest order approximation is already accurate, 
the intermediate one where the inclusion of the first correction
produces a good agreement with the
exact results and, finally, the one with smallest $\lambda$ where 
presumably the higher order correction of the flow is needed.
Some details of the specifics of the infrared regulator 
are also discussed.
\end{abstract} 

\date{\today} 
\maketitle


\section{Introduction}
In spite of the enormous success of functional renormalization group 
(FRG) approach
in statistical mechanics and quantum field theory (see 
\cite{DUPUIS20211} for a recent review), 
its applicability to strongly non-perturbative problems is not obvious.
In particular, the calculation of the energy gap $\Delta E$ between the 
first excited state and the ground state for the  anharmonic oscillator 
is an important test to study the effectiveness of the FRG approach 
to capture genuine topological effects. This model is based on the one 
particle  Hamiltonian with potential (here $\varphi$ indicates the
coordinate of the particle) :
\begin{equation}
\label{barev}
    V(\varphi)=\frac{M^2}{2}\;\varphi^2 +\lambda \varphi^4
\end{equation}
 which corresponds to the anharmonic oscillator with 
quartic corrections if $M^2>0$, or to the 
double well potential if $M^2<0$.
In the following, we express all dimensionful quantities 
in terms of the 
square mass scale $|M^2|$, which is equivalent to choose either 
$M^2=1$ or $M^2=-1$.

In the former case, a quartic term is added to the  exactly solvable
harmonic oscillator Hamiltonian, and therefore a perturbative treatment
of the problem is suitable as long as the quartic coupling does not 
grow too large.  In the latter case, the quadratic part is unstable and 
the stabilizing quartic term produces a double well that becomes deeper 
and deeper when $\lambda\to 0$. This means that the effect of the 
tunnelling of the wave function becomes more important for 
$\lambda\to 0$,
and therefore in this limit any perturbative approach fails to produce 
reasonable results.

A reliable approach to confront the $M^2=-1$,  $\lambda\to 0$  problem 
is the  dilute instanton gas calculation \cite{Coleman:1985rnk, Zinn-Justin:572813} 
that produces the well known, non-analytic exponential expression 
of the energy gap $\Delta E$ between first excited and ground state 
\begin{equation}
    \Delta E_{inst} = 2\; \Biggl ( \frac{2\sqrt{2}}{\pi\lambda}\Biggr)^{1/2}
  \; {\rm e}^{-\, {\left (3\lambda\sqrt{2}\right)^{-1}}}
\label{deinstanton}  
\end{equation}
Clearly, the quantum mechanical problem, either with $M^2=1$ or 
$M^2=-1$,
can be solved through the numerical determination  of the eigenvalues 
of the associated \scro equation which then provides the 
exact reference values of $\Delta E_{exact}$ at each value of $\lambda$.

Consequently, the calculation of $\Delta E$ represents an important
challenge for the FRG approach, in particular for low values of 
$\lambda$. 
Moreover, this analysis can be regarded as an essential step toward the 
application of the FRG to more complex problems regarding the 
estimate of false vacuum decay rates in quantum field
theory\cite{Devoto:2022qen} and, more specifically, concerning
the stability of the electroweak vacuum,
related to the Ultraviolet (UV) completion of the Standard Model, 
and the study of inflationary models in cosmology 
\cite{Bezrukov:2007ep},
because in these contexts the typical approach adopted mainly relies
on instanton calculation (see e.g.
\cite{Bentivegna:2017qry,Branchina:2019tyy}).

Early works started the investigation of this problem by means of 
a sharp cut-off FRG in the local potential approximation (LPA), either 
within a polynomial truncation
of the resulting flow equation for the potential
\cite{Horikoshi:1998sw, aoki}, or 
by direct numerical resolution of the local potential flow equation 
\cite{tetradis}. 
The inclusion of a field dependent wave-function renormalization 
$Z(\phi)$ first appeared  in \cite{Zappala:2001nv} by means of a 
Schwinger's proper-time  cutoff in the FRG equation. In that work a 
numerical 
integration of the combined system of non-linear parabolic Partial 
Differential Equations (PDE) for the local potential $V(\phi)$ and 
$Z(\phi)$ was employed.

In fact, polynomial truncations of the FRG generate a  system of 
differential equations which is singular  in the 
$k\rightarrow 0$  limit in the broken phase. On the contrary, the full 
set of PDE correctly reproduces the discontinuity in the correlation 
length below the critical temperature  \cite{parola} but the numerical 
integration is not straightforward,  as one has to resort to implicit 
methods for a class of coupled non-linear parabolic evolution equation  
\cite{Bonanno:2004pq,Caillol:2012zz}.
In  $d=1$ dimensions the evolution of flow in the infrared is less 
severe and one can hope that  
already with   the familiar {\it method of the lines} (MoL) \cite{ames1977numerical}, 
it is possible to reach values of $\lambda$ small enough
to render the comparison with standard dilute instanton gas approach meaningful. 
In \cite{Weyrauch:2006aj},  FRG  with smooth and sharp cut-off have been solved with 
both polynomial truncation and MoL but some discrepancies between the findings 
of \cite{Weyrauch:2006aj} and  \cite{Zappala:2001nv}  for very small $\lambda$ have emerged. 
Further applications of FRG to supersymmetric quantum mechanics have 
appeared in \cite{Synatschke:2008pv}. An attempt to apply the 
Principle of Minimal Sensitivity  in this context has recently been discussed in \cite{Kovacs:2014mia}.

Beyond LPA a number of structural aspects in the formulation of FRG arise.
In the case of the so called {\it spectrally adjusted} (SA) flow 
the coarse-graining procedure is built  in terms of the eigenvalues of the spectrum of $\Gamma_k^{(2)}$  evaluated at the  background field 
\cite{Gies:2002af}, where $\Gamma_k$
is the effective average action at the scale $k$. 
The spectrum is therefore not fixed but computed at the running cutoff $k$. 
If we visualize the averaging procedure in real space, the use of a
running cutoff built with a running blocked field can be thought as the analogous of a "lagrangian" description of the dynamic of the fluid made by a co-moving observer. 
On the contrary  the "eulerian"  coarse-grainined flow is produced if, 
in lowering the cutoff from $k$ to $k-\Delta k$,  a {\it fixed} 
spectrum of the Laplacian operator  $\Box$ is employed at each $k$.   
Clearly both  "lagrangian" and "eulerian"
schemes lead to very similar results near criticality  when the 
anomalous dimension is small,
but it is not clear which scheme is to be expected to work better in 
general. 

From this point of view we would like to stress that, strictly speaking,
only  FRG with {\it not spectrally adjusted} (NSA) regulators are
"exact" flow in the sense of \cite{Wetterich:1992yh,Morris:1993qb}, 
while SA regulators, albeit widely used in the FRG literature, are not 
in general 
exact unless they are derived from a  much more involved and non-linear
flow equation which we will not consider in this paper 
\cite{Litim:2002hj}.

The aim of this work is twofold. 
First, we would like to address the computation of $\Delta E$ 
in the small $\lambda$ regime, by means of distinct formulations,
namely the Exact Renormalization Group (ERG) flow equations 
characterized by the theta-function regulator proposed in 
\cite{Litim:2000ci} and the Schwinger's Proper Time (PT) flow 
equations introduced in \cite{bonannoazappala,Bonanno:2004sy,dealwis},
in order to get an inclusive prediction of $\Delta E$ from the FRG.
This point is essential both to test the reliability of this approach 
in the deep non-perturbative regime ($\lambda<<1$), especially 
in comparison with the instanton gas calculation,
and, at the same time, to produce a countercheck on the  mentioned
discrepancy with the analysis of \cite{Weyrauch:2006aj}.
Second, we intend to study the impact of SA {\it vs.} NSA cutoff in the 
low $\lambda$ region for these type of FRG. In particular,
in the case of the PT flow, we further discuss the dependence 
of the various cutoff schemes introduced in \cite{bonannolippoldt}.

The structure of the paper is the following. In Sect. \ref{flow} we
review the structure of the flow equations 
that are subsequently integrated and, more specifically, Sect.
\ref{ergfl} is devoted to the ERG 
flow,  while in Sect. \ref{ptfl} the PT  equations 
are discussed. Then, in Sect. \ref{numr} the results of our numerical 
analysis are presented, first considering the LPA approximation in
\ref{lpa}, and by including  the wave function renormalization
in \ref{zeta}. Our conclusions are reported in Sect. \ref{conc}.

\section{Flow equations\label{flow}}

\subsection{ERG flow \label{ergfl}}

We shall first consider the implementation of the ERG flow \cite{DUPUIS20211}
(we define the RG 'time'as $t=\log(k/\Lambda)$, where $k$ is the 
running energy scale and $\Lambda$ is a fixed UV cutoff)
\be
\label{erg}
\partial_{t} \Gamma_{k}[\varphi] = \frac{1}{2} {\rm Tr} \Bigl [ \partial_t{R}_k(q^2) \; G_k(q^2,\varphi) \Bigr ]
\ee
and the regularized propagator $ G_k(q^2,\varphi)$ is 
 \begin{equation}
 \label{propagator}
 G_k(q^2,\varphi)	=	\frac{1}{\Gamma^{(2)}_k(q^2,\varphi)+R_k(q^2)}
  =	\frac{1}{Z_k(\varphi)\, q^2+V''_k(\varphi)+R_k(q^2)} \;.
\end{equation}
$R_k(q^2)$ is the regulator of the infrared modes which will be 
specified below.
$\Gamma^{(2)}_k(q^2,\varphi)$ indicates the second functional 
derivative of  $\Gamma_{k}[\varphi]$
with respect to the field $\varphi$. Here and below, prime, 
double prime, etc. indicate one, two, etc. derivatives 
with respect to the  argument. 

The explicit form adopted for the running effective action 
$\Gamma_{k}[\varphi]$
comes from the  first non trivial order of its derivative expansion:
\be
\label{derexpa}
\Gamma_{k}[\varphi] = \int {\rm d}^dx \left[V_{k}(\varphi)+ \frac{1}{2}Z_{k}(\varphi)(\partial\varphi)^{2}\right] +O(\partial^{4})
\ee
and we focus on the  flows of $V_k$ and  $Z_k$, which are  derived from the flow  of the two-point function
\begin{eqnarray}
&\partial_{t}\Gamma_k^{(2)}(p^2) \;\;=\;\;  \partial_t Z_k\, p^{2}+\partial_tV''_k \;\;=\nonumber\\
&\int_{q}\left[ \Gamma_k^{(3)}(q,p,-q-p)\right]^2G_k \big((q+p)^2\big)G_k^2(q^2)\partial_t{R}_k(q^2) -\frac{1}{2}\int_{q}
\Gamma_k^{(4)}(q,p,-p,-q)G_k^2(q^2)\partial_t{R}_k(q^2)=\nonumber\\
&\int_{q}\left[Z'_k\left(q^{2}+q\cdot p+p^{2}\right)+V'''_k\right]^{2}G_k \big((q+p)^2\big)G_k^2(q^2)\partial_t{R}_k(q^2) \nonumber \\
&-\frac{1}{2}\int_{q}\left[Z''_k\left(q^{2}+p^{2}\right)+V''''_k\right]G_k^2(q^2)\partial_t{R}_k(q^2)
\label{gamma2flow}
\end{eqnarray}
by respectively extracting the $O(p^0)$ and $O(p^2)$ coefficients 
in the expansion of the left hand
 side (lhs) of Eq. (\ref{gamma2flow}) in powers of momentum $p$.
Incidentally, we notice that  in Eq. (\ref{gamma2flow}) 
the symmetry $\Gamma_k^{(3)}(q,p,-q-p)=\Gamma_k^{(3)}(q+p,-p,-q)$ is 
used to simplify the first contribution (diagram).

Therefore,  we need to expand the propagator $G_k \big({\tiny 
(q+p)^2}\big) $ in Eq. (\ref{gamma2flow}), in powers of $p$ 
\begin{equation}
\label{propagator_expansion}
G_k \big({\tiny (q+p)^2}\big) 
 =  G_k(q^{2})+2 x q \, G'_k(q^{2})p+\bigl 
 [G'_k(q^{2})+2x^{2}q^{2}G''_k(q^{2}) \bigr ]p^{2}+O(p^{3})
 )\,.
\end{equation}
where we recall that primes indicate derivative with 
respect to the argument  and we  labelled  the cosine of the angle 
between the four-vectors 
$p$ and $q$, as $x$. 

Then, by definition,  the derivatives of the regularized propagator are
\begin{equation}\label{propagator_derivative}
G'_k=-G^{2}_k(Z_k+R'_k) \qquad \qquad 
G''_k=2G^{3}_k(Z_k+R'_k)^{2}-G^{2}_kR''_k
\end{equation}
and  we can insert Eqs. (\ref{propagator_expansion}), 
(\ref{propagator_derivative})
into Eq. (\ref{gamma2flow}), to obtain the flow equations 
for $V_k$ and  $Z_k$
\begin{equation}
\label{ddimflow}
\partial_{t}V_{k}=\mathcal{B}_{V}^{d}\!\left(V_{k}'',Z_{k}\right)
\qquad \qquad 
\partial_{t}Z_{k}=
\mathcal{B}_{Z}^{d}\!\left(V_{k}'',V_{k}''',Z_{k},Z_{k}',Z_{k}''\right)
\end{equation}
where the explicit expressions for the right hand side (rhs) are
\begin{equation}
\label{ddimflowv}
(4\pi)^{\frac{d}{2}}\,\mathcal{B}_{V}^{d}=\frac{1}{2}
Q_{\frac{d}{2}}\!\left[G_{k}\partial_{t}R_{k}\right]
\end{equation}
and
\begin{eqnarray}
\label{ddimflowz}
(4\pi)^{\frac{d}{2}}\,\mathcal{B}_{Z}^{d}  \;\;\;\;=&
 \left(V_{k}'''\right)^{2} &\left\{ Q_{\frac{d}{2}}\left[G_{k}^{2}G_{k}'\partial_{t}R_{k}\right]+Q_{\frac{d}{2}+1}\left[G_{k}^{2}G_{k}''\partial_{t}R_{k}\right]\right\} 
 \nonumber\\
 & + Z_{k}'V_{k}''' & \left\{2Q_{\frac{d}{2}}\left[G_{k}^{3}\partial_{t}R_{k}\right]+(d+2)Q_{\frac{d}{2}+1}\left[G_{k}^{2}G_{k}'\partial_{t}R_{k}\right]\right.
 \nonumber\\
 &&\;\;\left.+(d+2)Q_{\frac{d}{2}+2}\left[G_{k}^{2}G_{k}''\partial_{t}R_{k}\right]\right\} 
 \nonumber\\
 & +\left(Z_{k}'\right)^{2} & \left\{ \frac{2d+1}{2}Q_{\frac{d}{2}+1}\left[G_{k}^{3}\partial_{t}R_{k}\right]
 +\frac{(d+2)(d+4)}{4}Q_{\frac{d}{2}+2}\left[G_{k}^{2}G_{k}'\partial_{t}R_{k}\right]\right.
 \nonumber\\
 &  & \;\;\left.+\frac{(d+2)(d+4)}{4}Q_{\frac{d}{2}+3}\left[G_{k}^{2}G_{k}''\partial_{t}R_{k}\right]\right\} 
 \nonumber\\
 & +\;Z_{k}'' & \left\{ -\frac{1}{2}Q_{\frac{d}{2}}\left[G_{k}^{2}\partial_{t}R_{k}\right]\right\}
\end{eqnarray}
and  the  $Q$--functionals of variable $z=q^2$ are defined as:
\begin{equation}
\label{qu}
Q_{n}[f] \equiv \frac{1}{\Gamma[n]} \int_0^\infty \,dz \, z^{n-1}f(z)
\end{equation}

So far, all relations are valid for an arbitrary cutoff function and in 
any dimension. 
Now, we specify the function $R_k(z)$ by taking
Litim's regulator \cite{Litim:2000ci}, in two different forms, 
namely in the plain NSA  version  
\begin{equation}\label{RkNSA}
R^{NSA}_k(z) = (k^{2}-z)\,\theta(k^{2}-z) 
\end{equation}
and in the SA form that includes 
a field dependent wave-function renormalization prefactor
\begin{equation}\label{RkSA}
R^{SA}_k(z) = Z_k(\varphi)\, (k^{2}-z)\,\theta(k^{2}-z) \,,
\end{equation}
and $\theta(x)$ is the  Heaviside function.

In the NSA case, only a partial cancellation of the square momentum 
$z$ dependence of $G_{k}(z)$ is realized, with the residual 
dependence weighted by the deviation of the factor $Z_k(\varphi)$ 
from 1 :
\begin{equation}
\label{gNSA}
G_{k}(z) = \frac{1}{Z_k \,z + (k^2-z) \theta(k^{2}-z) + V''_k} 
\end{equation}
and the replacement of Eqs. (\ref{RkNSA}), (\ref{gNSA}) into 
Eqs. (\ref{qu}), (\ref{ddimflowv}) and (\ref{ddimflowz})
leads us to a coupled pair of flow equations for $V_k$ and $Z_k$,
where the integration over the variable $z$ in Eq. (\ref{qu}) 
produces a long and 
involved sum of Hypergeometric functions, due to the not full momentum 
simplification in Eq. (\ref{gNSA}). 
Still, in this case, it is possible to 
recover an explicit, analytic form, suitable for numerical integration 
(which is discussed in the next Section),
of the partial differential equations (PDE) governing the flow of $V_k$ 
and $Z_k$, that, for the sake of simplicity, we do not display here.

The SA regulator displayed in (\ref{RkSA}) instead produces a much 
simpler structure of the flow equations, thanks to the presence of 
$Z_k(\varphi)$ in $R^{SA}_k(z)$. In fact, in this case the 
propagator reduces to the square  momentum $z$ independent quantity 
${H_k}^{-1}(\varphi)$ :
%
%
%
%
%
\begin{equation}
\label{gredux}
G_{k} (z) = \frac{1}{Z_k z + Z_k (k^2-z) \theta(k^{2}-z) + V''_k} 
\;\to\; \frac{1}{Z_k k^2+V''_k} \;\equiv\; \frac{1}{H_k}
\end{equation}
and the corresponding
$Q$--functionals entering $\mathcal{B}_V^d$ and $\mathcal{B}_Z^d$, 
have the following simple form:
\begin{equation}
\label{qu1}
Q_{n}[G_{k}^{m}\partial_{t}R_{k}]
= \frac{k^{2(n-m+1)}}{\Gamma(n)} \frac{Z_k}{(Z_k+\omega_k)^m}
\left\{\frac{2-\eta_k}{n}+ \frac{\eta_k}{n+1} \right\}
\end{equation}
\begin{equation}
\label{qu2}
Q_{n}[G_{k}^{m}G'_{k}\partial_{t}R_{k}] = 0
\end{equation}
\begin{eqnarray}
\label{qu3}
Q_{n}[G_{k}^{m}G_{k}''\partial_{t}R_{k}] &=& 
 - \frac{k^{2(n-m-2)}}{\Gamma(n)} \frac{Z_k^2}{(Z_k+\omega_k)^{m+2}}
\end{eqnarray}
where we used the specific value of Heaviside function at the origin, $\theta(0)=1/2$, and 
we defined $\omega_k \equiv V''_k/k^2$
and  the field and scale dependent $\eta_k(\varphi)$
(recall $t=\log(k/\Lambda)$) as
\begin{equation}
\label{eta}
\eta_k(\varphi) \equiv -  \partial_t \log Z_k (\varphi) \;,
\end{equation}
that must not be mistaken for the field anomalous 
dimension at criticality.
Then for the SA case, 
 Eqs. (\ref{RkSA}), (\ref{qu1}), (\ref{qu2}), (\ref{qu3}),
and (\ref{gredux}) yield the following 
 $\mathcal{B}$-functions  in $d=1$, to be inserted in 
Eq. (\ref{ddimflowv}) 
\begin{equation}
\label{saERGV}
\mathcal{B}_{V}^{d=1} =\frac{ Z_k k^3}{\pi \,H_k} 
\;\left( 1-\eta_k /3 \right )
\end{equation}
\begin{eqnarray}
\label{saERGZ}
\mathcal{B}_{Z}^{d=1} = \;\frac{ Z_k k^3}{\pi \,H_k} &\cdot&
\Bigg\{ \left (-\frac{Z''_k  }{H_k}   
\,+\,  \frac{4\,Z'_k  V'''_k}{H_k^2}  \,\right ) 
 \left( 1-\frac{\eta_k}{3}\right ) \,+\,
\frac{2\,(Z'_k)^2 \, k^2 }{H_k^2 }  
\, \left( 1-\frac{\eta_k}{5}\right )
\nonumber\\
\,&-&\, \frac{Z_k  (V'''_k)^2   }{H_k^3} \,-\,
\frac{2 k^2\,Z_k Z'_k  V'''_k  }{H_k^3}  \,-\,
\frac{Z_k\,(Z'_k)^2 \, k^4   }{H_k^3}  \Bigg \}
\end{eqnarray}

Despite the compact form of the flow equations displayed in 
Eqs. (\ref{saERGV}) and (\ref{saERGZ}), we notice that 
terms proportional to $\eta_k$ introduce nonlinear effects 
related to the 'time' derivative of $Z_k$, which are harmless 
as long as $Z_k\simeq 1$, but become very difficult to 
handle in the numerical integration when $Z_k$ becomes 
substantially different from 1, which occurs in the region 
of very small coupling $\lambda$ in Eq. (\ref{barev}).
Due to this complication, together with the full flow of $V_k$ and 
$Z_k$ in Eqs. (\ref{saERGV}) and 
(\ref{saERGZ}), we shall analyze 
its reduced version obtained by  simply setting $\eta_k=0$, thus 
making the flow equations linear both in 
$\partial_t V_k$ and
$\partial_t Z_k$, and therefore much easier to integrate numerically.

\subsection{Proper Time flow \label{ptfl}}
Now, we turn to a different kind of flow, namely the Proper Time (PT) 
flow, whose equations for $ V_k$ and $ Z_k$ can be cast in the 
form of PDE, suitable for numerical investigation. 
This kind of flow is to be regarded as a particular case of
background field flow \cite{litimpawlow}, although recently 
it was reconsidered as a type of Wilsonian action flow
\cite{dealwis,bonannolippoldt},
and the corresponding flow  equations of 
$ V_k$ and $ Z_k$ are discussed in detail in \cite{bonannolippoldt}.  
Here, we do not focus on the nature of the PT flow, as we are rather 
interested in its application to the spectrum of the double well 
potential and we  refer to  \cite{bonannolippoldt}
for a detailed discussion on the structure of the regulator  
and on the consequent derivation of the flow
equations, both in the NSA and in the SA case. 

Here, we just mention that  the NSA regulator, which carries no 
dependence on the renormalization factor $Z_k$, produces the 
following flow equation  (named "A-scheme" in \cite{bonannolippoldt}) 
\begin{equation} 
\label{nsaPT}
\partial_{t} \Gamma_{k}[\varphi] = - {\rm Tr} \left( \frac{m 
k^{2}}{\Gamma^{(2)}_{k}[\varphi] + m k^{2}} \right)^{m} .
\end{equation}
where $m>d/2$ is a free parameter that 
roughly specifies the sharpness of the regulator, 
and it is usually  taken as integer.
Eq. (\ref{nsaPT}) in $d=1$ and with the same formal parameterization 
of the action adopted in Eq. (\ref{derexpa}), 
reduces to the following NSA coupled flow 

\begin{equation}
\label{nsaPTV}
\partial_{t} V_{k}=- \; \frac{\Gamma(m-1/2)}{ 2\; \sqrt {\pi} 
\;\Gamma(m) }
\;\,\sqrt{\frac{m\,k^2}{Z_k}}\;\; \left ( \frac{ \;m \;k^2 
\;}{\;\;h^{^{NSA}}_{k,\,m}\;\;} \right )^{m-1/2}
\end{equation}
\begin{eqnarray}
\label{nsaPTZ}
\partial_{t} Z_{k}&=&\;  \frac{ \Gamma(m+1/2)}{ 2\; \sqrt {\pi} 
\;\Gamma(m) }
\;\,\sqrt{\frac{m\,k^2}{Z_k}}\;\; \left ( \frac{ \;m \;k^2 
\;}{\;\;h^{^{NSA}}_{k,\,m}\;\;} \right )^{m-1/2}\cdot\nonumber\\
&&\left [\; \frac{Z''_k}{h^{^{NSA}}_{k,\,m}  } 
+ \frac{Z_k  (V'''_k)^2 
(m+1/2)(m+3/2)  }{6 \left (h^{^{NSA}}_{k,\,m}\right )^3}  
- \frac{3\,Z'_k 
V'''_k  (m+1/2) }{ 2 
\left (h^{^{NSA}}_{k,\,m}\right )^2}   
\,-\, \frac{21}{24}\,\frac{(Z'_k)^2
}{Z_k \, h^{^{NSA}}_{k,\,m}  }   \right ]    
\end{eqnarray}
where $\Gamma(x)$ is the Gamma-function and $h^{^{NSA}}_{k,m}$ is 
\begin{equation}
\label{acca}
h^{^{NSA}}_{k,\,m}=  m \, k^2 + V_k''
\end{equation}

On the other hand, the SA flow requires the replacement of the 
scale $k^2$  with the corrected scale $ Z_k(\varphi) \;k^2$ in the 
regulator and this produces flow equations indicated in
\cite{bonannolippoldt} as the 'B-scheme', or
as the 'simplified type-C scheme' if the derivatives of the 
factors $ Z_k(\varphi)$ that appear in the regulator are neglected. 
We refer to  \cite{bonannolippoldt} for the explicit involved form
of the flow equations  of $ V_k$ and $ Z_k$ in the 'B-scheme', while 
the 'simplified type-C scheme' equations are simply obtained from
 Eqs. (\ref{nsaPTV}) and (\ref{nsaPTZ}) by replacing everywhere
$m \;k^2 \;\to\; Z_k(\varphi)  \, m \;k^2 $, i.e. we find 
\begin{equation}
\label{saPTV}
\partial_{t} V_{k}=\,-  \; \frac{\Gamma(m-1/2)}{ 2\; \sqrt {\pi} 
\;\Gamma(m) }       \;\,\sqrt{m\,k^2}\;\; \left ( \frac{ Z_k \;m \;k^2 
\;}{\;\;h^{^{SA}}_{k,\,m}\;\;} \right )^{m-1/2}
\end{equation}
\begin{eqnarray}
\label{saPTZ}
\partial_{t} Z_{k}&=&
\;  \frac{ \Gamma(m+1/2)}{ 2\; \sqrt {\pi} \;\Gamma(m) }
 \;\,\sqrt{m\,k^2}\;\; \left ( \frac{  Z_k\;m \;k^2 
 \;}{\;\;h^{^{SA}}_{k,\,m}\;\;} \right )^{m-1/2}\cdot\nonumber\\
&&\left [\; \frac{Z''_k}{h^{^{SA}}_{k,\,m}  } + \frac{Z_k  (V'''_k)^2 
(m+1/2)(m+3/2)  }{6 \left (h^{^{SA}}_{k,\,m}\right )^3}  
- \frac{3\,Z'_k  V'''_k  (m+1/2) }{ 2 
\left (h^{^{SA}}_{k,\,m}\right )^2}   
\,-\, \frac{21}{24}\,\frac{(Z'_k)^2  }{Z_k \, h^{^{SA}}_{k,\,m}  }
\right ]    
\end{eqnarray}
with the definition 
 \begin{equation}
\label{accasa}
h^{^{SA}}_{k,\,m} \equiv  Z_k\; m \, k^2 + V_k'' \;.
\end{equation} 

We notice that the 'simplified type-C scheme' flow equations
(\ref{saPTV}) and (\ref{saPTZ}) 
are equal to   the pair of equations analyzed in 
\cite{Zappala:2001nv} (and previously introduced in 
\cite{bonannoazappala}), provided that the scale $\hat k$ and the 
parameter $\hat m$, used in 
\cite{Zappala:2001nv}, are replaced with our $k$ and $m$, 
according to $\hat m +1= m$ and $\hat k =k \, m/ \hat m$.  
Clearly, observables calculated 
in the limit  $k\to 0$ are insensible to the difference between $k$ 
and $\hat k$, and one can compare the results 
derived from the two flows for equivalent values of  $m$ and $\hat m$.

Computations in \cite{Zappala:2001nv} 
were performed with a particular exponential
form of the flow obtained  in the limit $\hat m \to \infty$. 
In fact, the same exponential flow equations are recovered from Eqs. 
(\ref{saPTV}) and (\ref{saPTZ}) in the limit $m\to \infty$; however we 
shall not repeat 
here the computation of $\Delta E$ in this case, but rather look at 
 the PT flow 
at the reasonably large value $m=10$, to
produce predictions close to those  of the exponential flow.

In addition, we notice that the SA version of the PT flow in  
Eqs. (\ref{saPTV}) and (\ref{saPTZ})  at the particular value  $m=3/2$, 
resembles the ERG  displayed in Eqs. (\ref{saERGV}) and 
(\ref{saERGZ}), at $\eta_k=0$. 
Actually, for $m=3/2$, the flow equations for $V_k$ (\ref{saPTV})
and (\ref{saERGV}) coincide, while in  the  equations for $Z_k$,  
only some terms are identical. From the numerical analysis, it will be 
evident that the differences between the two are subleading and the 
ERG computation of $\Delta E$ falls between the values of $\Delta E$ 
determined with the PT flow at $m=1$ and $m=2$.

To summarize, in the next Section we integrate both 
NSA and SA version of the PT flow, respectively in  
Eqs. (\ref{nsaPTV}) and (\ref{nsaPTZ}) and 
in Eqs. (\ref{saPTV}) and (\ref{saPTZ}) at small values of $m$, namely 
at $m=1,2,3,4$, at which the convergence of the differential equations 
is more accurate, 
and also at $m=10$ which, as already noticed, should produce results 
that are close to those obtained with the exponential PT flow 
\cite{Zappala:2001nv}.


\section{Numerical Results\label{numr}}

\subsection{Preliminary details \label{prel} }

\begin{table}[htb]
\begin{center}
\begin{tabular}
{|c  | c | c |  c|}
 \hline \hline\hline
$M^2=1$&&& \\
$\;\;\;\lambda\;\;\;\;$ & $\;\;\;\Delta E_{exact}\;\;\;$&  $-$  &$\;\;\;\Delta E_{WH}\;\;\;$ \\
\hline\hline
 1      &1.9341   & $-$ & 1.928(2)   \\
\hline
 0.02 &1.0540   & $-$ & 1.053(2)   \\
\hline\hline\hline\hline 
$M^2=-1$&&&\\
$\;\;\;\lambda\;\;\;\;$ & $\;\;\Delta E_{exact}\;\;$&$\;\;\Delta E_{inst}\;\;$ &$\;\;\Delta E_{WH}\;\;$ \\
\hline\hline
\small 0.4   & 0.9667   &  1.6645 & 0.965(2) \\
\hline
0.3    & 0.8166   &  1.5792 & 0.817(2) \\
\hline
0.2    & 0.6159   &  1.3058 & 0.621(2) \\
\hline
0.1    & 0.2969   & 0.5683  & 0.329(2)  \\
\hline
0.05  & 0.0562 & 0.0761   &  0.157(2)  \\
\hline
0.04  & 0.0210 & 0.0262 & 0.125(2) \\ 
\hline
0.03  & 0.0036 &  0.0042 & 0.094(1)  \\ 
\hline
0.02  &  0.93\;$10^{-4}$  &  1.02\;$10^{-4}$  & 0.063(1) \\
\hline
0.01  &  1.05\;$10^{-9} $ &1.10\;$10^{-9}$ & 0.032(1)  \\
\hline\hline
\end{tabular}
\end{center}
\caption{$\Delta E$ for various couplings $\lambda$ and for $M^2=1$ or 
$M^2=-1$, as obtained from
the Schroedinger equation ($\Delta E_{exact}$), the instaton 
calculation ($\Delta E_{inst}$), and from the Wegner-Houghton flow 
equation ($\Delta E_{WH}$). }
\label{tab0}
\end{table}  

As discussed previously,
the most accurate determination of the 
energy gap $\Delta E$ comes from the resolution 
of the eigenvalue problem 
for the associated Schroedinger equation. In the following,
we indicate with $\Delta E_{exact}$ the gap computed in this way and
display in Table \ref{tab0} some determinations both for 
$M^2=1$ and $M^2=-1$. In the former case, only the two values $\lambda=1$
and $\lambda=0.02$ are selected to test the predictions of the RG flow
either in a strongly or in a weakly coupled  regime.
Conversely in the case $M^2=-1$, where the non-perturbative effect of 
the tunnelling becomes important, we select more values of the quartic 
coupling in the range $\lambda=0.4$ to $\lambda=0.01$, to have a clear 
understanding of the RG flow predictions in different regimes.

In Table \ref{tab0}, for the double well potential case $M^2=-1$
we include the data of the gap as obtained from the instanton 
calculation reported in Eq. (\ref{deinstanton}).  
It is immediately evident  that for  $\lambda=0.04$ or larger,
$\Delta E_{inst}$  are very far form $\Delta E_{exact}$, and even  at
$\lambda=0.03$ we find an error of about $30\%$. Conversely,
the data reported at  $\lambda=0.02, \, 0.01$ show respectively 
an error of $10\%$ and  $5\%$. Only for lower values of $\lambda$
the instanton calculation becomes really precise.

Now, we are interested in obtaining 
a further determination of $\Delta E$ 
from the numerical resolution of the flow equations for 
$V_k$ and $Z_k$ down to $k=0$. In fact, $(\Delta E)^2$ is related 
to the renormalized curvature of  the effective potential at the 
origin $({V_{k=0}(\varphi)''}/{Z_{k=0}(\varphi)})\;\big|_{\varphi=0}$.
However, the resolution of the flow equations when the
initial condition (set at a large UV scale $k=\Lambda$)
is given by the  potential in (\ref{barev}) with $M^2=-1$
(which is  concave downward around $\phi=0$),
usually presents a stiff behavior related to the appearance of a
spinodal line. This line is defined by the vanishing of the 
inverse propagator entering the flow equations: i.e. in our case 
it is defined by the vanishing 
of $G_k(z)^{-1}$ in Eq. (\ref{gNSA}) or of $H_k$, $h^{NSA}_k$,
$h^{SA}_k$, respectively defined  in Eqs. (\ref{gredux}), (\ref{acca}),
(\ref{accasa}).

This problem is well known since long time. A systematic way of 
circumventing  was proposed in \cite{parola}, and further developed 
in \cite{Bonanno:2004pq} and \cite{Caillol:2012zz}. Essentially, it consists in 
computing the flow of a novel variable, defined as a particular 
function of the inverse propagator chosen in accordance to the 
structure of the original flow equations; so, for instance, 
the form of Eq. (\ref{nsaPT}) suggests the introduction of the 
new variable $(h^{NSA}_k)^{1/2-m}$ and the replacement
of the flow equation of $V_k$ with the one for 
this new variable. 

Typically, the flow equation for this new variable,
coupled with the one for $Z_k$, 
has a stabler behavior in proximity of the spinodal  line. 
Then, it is convenient to take advantage of this alternative
approach and therefore we checked the consistency of 
all our determinations, by comparing  the output of the flow of 
the original variables and of the new ones.

Concerning the numerical methods adopted, we compared 
different approaches for each single determination of $\Delta E$.
Specifically, in one case we integrated the system of PDE
by performing the  spatial discretisation with  a Chebyshev 
collocation method, and employing the method of lines to reduce 
the PDE to a system of ordinary differential equations. 
The resulting system is solved using a backward 
differentiation formula method, as encoded in the numerical 
libraries NAG \cite{NAG}.

As an alternative method, we have followed the approach of 
\cite{parola,Bonanno:2004pq} and we have used the MoL
on a set of transformed equations for the threshold functions 
\cite{Caillol:2012zz}. For actual calculations in this case
we have employed the MoL as in \texttt{Mathematica} \cite{Mathematica:12.1}
with  \texttt{AccuracyGoal=10} and  \texttt{PrecisionGoal=10}.

The first attempt to determine $\Delta E$ from the RG flow was 
obtained by solving the Wegner-Houghton equation \cite{Horikoshi:1998sw,
aoki,tetradis}
\begin{equation}
\label{WH}
\partial_{t} V_{k}= \; \frac{k}{ 2\; \pi} 
\; {\rm log}\, \Bigl (1+ \frac{V_k(\varphi)''}{k^2} \Bigr )\;.
\end{equation}

It is a single equation for the running potential $V_k$, and
its generalization to include a renormalization factor $Z_k$ 
is usually not taken into account as it contains some ambiguities 
\cite{bonannozappalaprd,bonannobranchina}.
Then, by integrating   Eq. (\ref{WH}),
one finds a convex potential at $k=0$, which is already a remarkable 
feature, because it is an exact property, \cite{Symanzik:1969ek}, 
that can be recovered
only within  non-perturbative approaches (see, e.g., 
\cite{Callaway:1982us, Callaway:1982si, Branchina:1990ja}) 
and, in addition, the gap  is  directly read from the value of the 
second derivative of the potential at the origin 
$\Delta E_{WH}=\sqrt{V_{k=0}(\varphi)''}|_{\varphi=0}$.

As a first exercise, we integrate Eq. (\ref{WH}) and the output
is reported in the last column of  Table \ref{tab0}.
As a consequence of the functional form of Eq. (\ref{WH}),
the results are not extremely accurate and $\Delta E_{WH}$
in Table  \ref{tab0} are typically affected by
an error of one or two units on the last displayed digit.
We notice that when  $M^2=1$, $\Delta E_{WH}$ 
differs from $\Delta E_{exact}$  of about $0.3\%$ for $\lambda=1$
and $0.1\%$ for $\lambda=0.02$ which shows the high accuracy of these 
findings in both regimes.
The trend of $\Delta E_{WH}$ when $M^2=- 1$, will be discussed in 
Sect. \ref{lpa}, in comparison with other estimates.

\subsection{Local Potential Approximation\label{lpa}}

\begin{table}[htb]
\begin{center}
\begin{tabular}{|c | c | c | c | c | c | c | c |}\hline\hline
$M^2=1$&&&&&&&\\
$\;\;\;\lambda\;\;\;\;$&$\;\;\Delta E_{exact}\;\;$&$\;\;\Delta E_{erg}\;\;$&
$\;\;\Delta E_{pt1}\;\;$  & $\;\;\Delta E_{pt2}\;\;$&  $\;\;\Delta E_{pt3}\;\;$& $\;\;\Delta E_{pt4}\;\;$&   $\;\;\Delta E_{pt10}\;\;$\\
\hline\hline
 1      &1.9341      &   1.9391 & 1.9361 &  1.9409 &  1.9428 & 1.9438 & 1.9458  \\
\hline
 0.02 &1.0540      &  1.0542  & 1.0541  & 1.0542 & 1.0543 &1.0543 & 1.0544  \\
\hline\hline\hline\hline 
$M^2=-1$&&&&&&&\\
$\;\;\;\lambda\;\;\;\;$ & $\;\;\Delta E_{exact}\;\;$&$\;\;\Delta E_{erg}\;\;$& 
$\;\;\Delta E_{pt1}\;\;$  & $\;\;\Delta E_{pt2}\;\;$&  $\;\;\Delta E_{pt3}\;\;$& $\;\;\Delta E_{pt4}\;\;$&   $\;\;\Delta E_{pt10}\;\;$\\
\hline\hline
0.4    & 0.9667    &  0.9785  & 0.9743  &  0.9810 & 0.9837   &  0.9852 & 0.9879 \\
\hline
0.3    & 0.8166     &  0.8295  & 0.8254 & 0.8319  & 0.8345 & 0.8359 &  0.8386  \\
\hline
0.2    & 0.6159     &  0.6314  & 0.6277 & 0.6336  & 0.6360 & 0.6373 & 0.6399  \\
\hline
0.1    & 0.2969     &  0.3243    &  0.3241 & 0.3248  & 0.3256  & 0.3261 & 0.3272 \\
\hline
0.05  & 0.0562     & 0.1120    &  0.1252 & 0.1049 & 0.0973 & 0.0932 & 0.0856   \\
\hline
0.04  & 0.0210     &  0.0796 & 0.0939  & 0.0717  & 0.0635  & 0.0595 &  0.0532  \\
\hline
0.03  & 0.0036    & 0.0545 & 0.0668  & 0.0481 & 0.0422 & 0.0395 & 0.0353 \\  
\hline
0.02  &  0.93\;$10^{-4}$   & 0.0341  & 0.0427  &  0.0299 & 0.0262   & 0.0246&  0.0219 \\
\hline
0.01  &  1.05\;$10^{-9} $   & 0.0161 & 0.0206  & 0.0141  & 0.0124  & 0.0116& 0.0104 \\
\hline\hline
\end{tabular}
\end{center}
\caption{$\Delta E$ for various couplings $\lambda$ and for $M^2=1$ or 
$M^2=-1$, as obtained from
the Schroedinger equation ($\Delta E_{exact}$), the LPA ERG flow 
($\Delta E_{erg}$)  
and  from the LPA PT flow 
with parameter $m=1,2,3,4,10$ ($\Delta E_{pt1-pt10}$). }
\label{tab1}
\end{table}  

In Table \ref{tab1} we report the numerical results obtained in the LPA,
i.e. by keeping the renormalization fixed to $Z_k=1$ along the flow,
in the various cases corresponding to the ERG flow in 
Eqs. (\ref{ddimflow}), (\ref{saERGV}) and to the PT flow in 
Eq. (\ref{nsaPTV}) for $m=1,2,3,4,10$. With no correction due to the 
normalization factor $Z_k$, the  gap in these computations is obtained 
as
\begin{equation}
\label{gaplpa}
   \Delta E =\sqrt{V_{k=0}(\varphi)''}|_{\varphi=0} \;\;.
\end{equation}
To facilitate the comparison, 
$\Delta E_{exact}$, already shown in Table \ref{tab0}, 
is again reported in in Table \ref{tab1}.
We notice that, as expected, $\Delta E_{erg}$
always sits between the corresponding $\Delta E_{pt1}$ and 
$\Delta E_{pt2}$.

\begin{figure}
\begin{centering}
\includegraphics[width=11 cm,height=8cm]{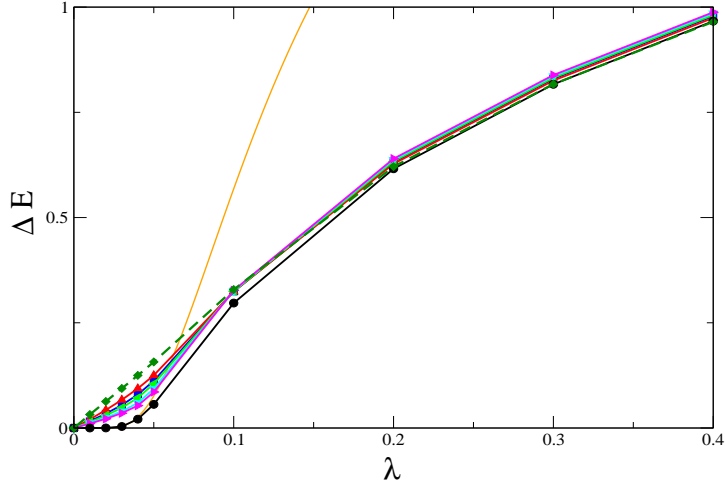}
\par\end{centering}
\caption{\label{figure1} Data reported in Tables \ref{tab0},  \ref{tab1}  
for $\Delta E$ as obtained in LPA, plotted as a function of the coupling $\lambda$. 
Lines joining the single points are included for convenience.  $\Delta E_{exact}$ and $\Delta E_{WH}$
from  Table \ref{tab0} are respectively displayed as black circles with a continuous black line and green diamonds 
with  a dashed  green  line. Then data from  Table \ref{tab1} are : blue squares, $\Delta E_{erg}$;  red pointing up triangles, $\Delta E_{pt1}$;
light green  pointing left triangles, $\Delta E_{pt2}$; turquoise pointing down triangles, $\Delta E_{pt4}$; pink  pointing right triangles, $\Delta E_{pt10}$,
all joined by continuous lines of the respective colour. The continuous orange line joins the instanton data of Table \ref{tab0}.}
\end{figure}

\begin{figure}
\begin{centering}
\includegraphics[width=13 cm,height=8cm]{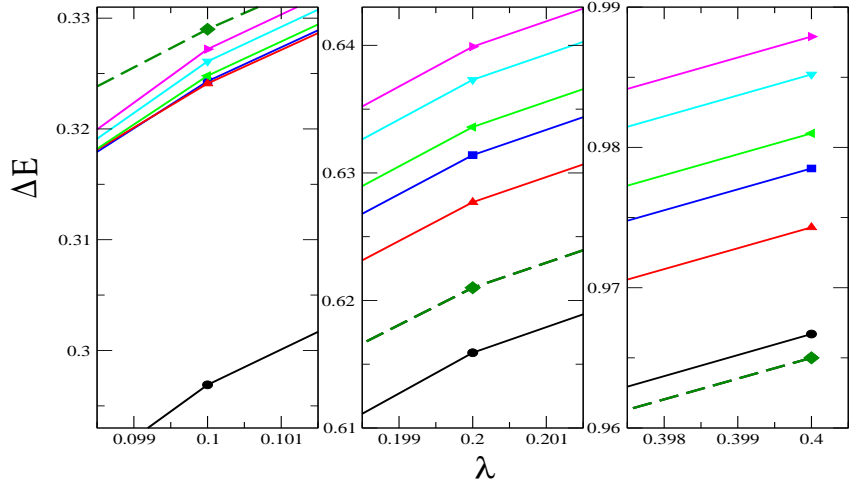}
\par\end{centering}
\caption{\label{figure2} Details of Fig.\ref{figure1} at three 
values  of the coupling, $\lambda=0.1,\, 0.2$ and  $0.4$. Coding is 
given in Fig.\ref{figure1}. }
\end{figure}

\begin{figure}
\begin{centering}
\includegraphics[width=11 cm,height=8cm]{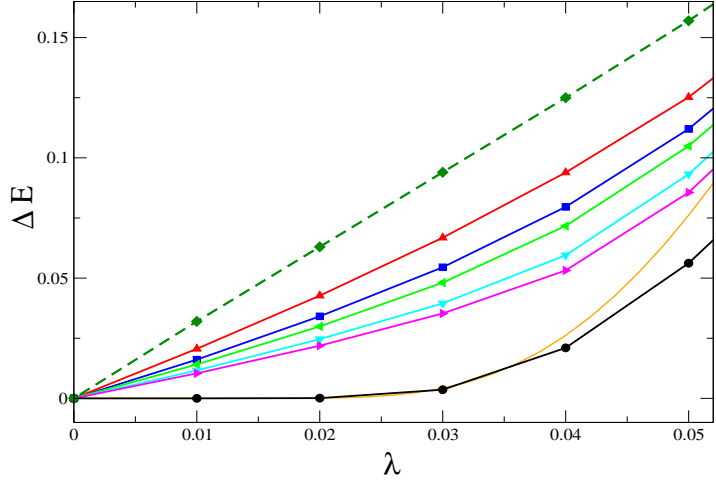}
\par\end{centering}
\caption{\label{figure3} Details of Fig.\ref{figure1} at small values of the coupling  $\lambda$. Coding is given in Fig.\ref{figure1}. }
\end{figure}

Then, the content of Tables \ref{tab0} and  \ref{tab1} is plotted in
Fig. \ref{figure1}, and the points corresponding to various columns
of these Tables are plotted with different colours and symbols, 
according to the  legend reported in the Figure caption (note that,
to avoid a redundant superposition of curves, in all figures we omit 
the data corresponding to $\Delta E_{pt3}$). 
Fig. \ref{figure2} shows three details of Fig. \ref{figure1} for 
$\lambda=0.4$,  $\lambda=0.2$ and $\lambda=0.1$, while 
Fig. \ref{figure3} is the enlargement of Fig. \ref{figure1} in the 
region  of smaller $\lambda$.

From Fig. \ref{figure1} it is evident that all computations
in the LPA produce accurate results only above $\lambda=0.1$.
In fact from the data in Table  \ref{tab1} we see that for $\lambda\geq
0.2$ the  maximum distance of the various determinations 
from $\Delta E_{exact}$ is below $4\%$ but it reaches $10\%$ at
 $\lambda=0.1$.

In Fig. \ref{figure2} it is shown that at $\lambda=0.4$ and 
$\lambda=0.2$,
$\Delta E_{WH}$ is the closest estimate to $\Delta E_{exact}$, while 
at $\lambda=0.1$ it becomes the farthest. At the same 
time, the distance from $\Delta E_{exact}$
of the PT estimates at various $m$
grows with $m$, when $\lambda\geq 0.1$.

A totally different picture appears in Fig. \ref{figure3} for 
$\lambda\leq 0.05$. In fact we observe that $\Delta E_{WH}$ 
approaches zero linearly with $\lambda$, then totally missing the 
exponential behavior of $\Delta E_{exact}$.
On the other hand, the LPA PT determinations invert their order and 
in this region the determinations with lager $m$ show better 
agreement with $\Delta E_{exact}$. In addition they also 
show a slight change of concavity that improves for larger $m$.

In summary, while our findings show that the LPA calculations are 
altogether satisfactorily accurate for $\lambda\geq 0.1$,
on the contrary they are quantitatively inadequate to reproduce 
$\Delta E_{exact}$ when $\lambda\leq 0.05$. In this case, some 
improvement is necessary.

\subsection{Inclusion of the renormalization $Z_k$\label{zeta}}

In this section we finally present the gap $\Delta E$,
corrected by the renormalization factor  $Z_k$, i.e.
\begin{equation}
  \label{gapfin}
   \Delta E =\sqrt{\frac{V_{k=0}(\varphi)''}{Z_{k=0}(\varphi)}}
  \; \;\Bigg|_{\varphi=0}  
\end{equation}
as obtained from the coupled flow equations for $V_{k}$ and $Z_{k}$.
Table \ref{tab2} contains the results of the  NSA ERG flow and of the
NSA PT flow
for $m=1,2,3,4,10$, together with the corresponding value of the 
correction  $Z_{k=0}(\varphi=0)$ displayed in parenthesis below.
Again, the  column  with $\Delta E_{exact}$ is inserted for 
comparison.

\begin{table}[htb]
\begin{center}
\begin{tabular}{|c | c | c | c | c | c | c | c |}\hline\hline
$M^2=1$ &&&&&&&\\
$\;\;\lambda\;\;$ & $\;\;\Delta E_{exact}\;\;$
&$\;\;\Delta E_{erg}\;\;$& $\;\;\Delta E_{pt1}\;\;$  & $\;\;\Delta E_{pt2}\;\;$&  $\;\;\Delta E_{pt3}\;\;$& $\;\;\Delta E_{pt4}\;\;$&   $\;\;\Delta E_{pt10}\;\;$\\
\hline\hline
 1   &1.9341 & 1.9351  & 1.9344    &  1.9365 &1.9384 &  1.9395 & 1.9417  \\
          && (1.0073) & (1.010)   &  (1.007) & (1.007)& (1.007)&  (1.006)  \\
\hline
0.02 &1.0540 & 1.5043& 1.0542    &1.0542     &1.0543  & 1.0543    & 1.0543 \\
                    && (0.9993) & (1.0003)  &(1.0003)  &(1.0002)& (1.0002) & (1.0002) \\
\hline\hline\hline\hline 
$M^2=-1$&&&&&&&\\
$\;\;\lambda\;\;$ & $\;\;\Delta E_{exact}\;\;$
&$\;\;\Delta E_{erg}\;\;$& 
$\;\;\Delta E_{pt1}\;\;$  & $\;\;\Delta E_{pt2}\;\;$&  $\;\;\Delta E_{pt3}\;\;$& $\;\;\Delta E_{pt4}\;\;$&   $\;\;\Delta E_{pt10}\;\;$\\
\hline\hline
0.4   & 0.9667  & 0.9764& 0.9632  & 0.9688 & 0.9723  & 0.9743  & 0.9782 \\
                       &&(1.037)& (1.043) & (1.033) & (1.029) &(1.027)  &(1.023)\\
\hline
0.3   & 0.8166  & 0.8276&  0.8120 & 0.8182  & 0.8219  &0.8239 &  0.8280 \\
                       &&(1.047)&  (1.055)& (1.042) & (1.037) &(1.034)  & (1.029)\\
\hline
0.2  & 0.6159   & 0.6300&  0.6083 & 0.6162 &  0.6203 & 0.6227&  0.6271\\
                      && (1.074)&  (1.088) & (1.065)& (1.056)  & (1.052)& (1.044)\\
\hline
0.1  & 0.2969   & 0.3158 & 0.2702 & 0.2888 & 0.2950 & 0.2980 & 0.3037 \\
                       &&(1.264)& (1.356)& (1.225) & (1.188)& (1.170) & (1.140)\\
\hline
0.05  & 0.0562 & X &  X & 0.0168 & 0.0262 & 0.0293 & 0.0333 \\ 
                       &&&    & (7.503)& (4.318)& (3.648) & (2.852)\\ 
\hline\hline
\end{tabular}
\end{center}
\caption{$\Delta E$ for various couplings $\lambda$ and for $M^2=1$ or 
$M^2=-1$, as obtained from
the Schroedinger equation ($\Delta E_{exact}$) and 
from the coupled flow  equation for  $V_k$ and $Z_k$,
respectively in the NSA ERG case
($\Delta E_{erg}$),  and the NSA PT case
with  $m=1,2,3,4,10$ ($\Delta E_{pt1-pt10}$). 
The correction $Z_k(\varphi=0)$ 
 is reported in parenthesis. }
\label{tab2} 
\end{table}  

The NSA PT determinations of $\Delta E$
when $M^2=1$ strongly improve the agreement 
obtained with the LPA calculation. In fact, from Table \ref{tab2} 
we observe  that at 
$\lambda=1$ the error for the worse determination 
($\Delta E_{pt10}$) is about $0.4\%$, but decreases to $0.02\%$ for 
$\Delta E_{pt1}$; at $\lambda=0.02$ the error is always around
$0.03\%$. In addition, for both values of $\lambda$, we find 
$\Delta E_{pt1} \leq \Delta E_{erg} \leq \Delta E_{pt3}$ as 
in the LPA case. In all determinations,
the correction factor $Z_{k=0}$ is extremely  small.

By switching to the case $M^2=-1$,  for the PT case
we find $\Delta E_{pt1}<\Delta E_{exact}<\Delta E_{pt10}$
for $\lambda>0.1$,
with an interval $(\Delta E_{pt10}-\Delta E_{pt1})/\Delta E_{pt10}$ 
of order 
$2\%\div 3\%$ if  $\lambda>0.2$, and around $10\%$ at $\lambda=0.1$.
Instead, at $\lambda=0.05$, we observe a rather different picture,
because  in the $m=1$ case an imaginary component of the potential
is generated and no clear convergence is observed whereas, at larger $m$,
a definite value of $\Delta E$  
together with very large $Z_{k=0} >>1$ is obtained
but these estimates substantially differ
from $\Delta E_{exact}$. 

For the NSA ERG data, we observe an unusual picture: in fact 
$\Delta E_{erg}$ is no longer similar  to  $\Delta E_{pt1}$ 
but it is rather closer to  $\Delta E_{pt10}$ and, below $\lambda=0.3$,
it becomes even bigger: $\Delta E_{erg} > \Delta E_{pt10}$.  
At $\lambda=0.05$, we do not find convergence to a definite value for
$\Delta E_{erg}$, much in the 
same way as it is observed  for $\Delta E_{pt1}$.

Finally, at smaller $\lambda$, no convergence
is found, neither for the NSA PT case at any $m$, nor for the
NSA ERG and we conclude that
if $M^2=1$ the NSA determinations of the gap improve the LPA results, 
but if $M^2=-1$, this is true only in the perturbative region 
($\lambda>0.1$ and $Z_{k=0} \sim 1$). At smaller $\lambda$ when 
the non-perturbative tunnelling effects become relevant,
computations become impracticable.

\begin{figure}
\begin{centering}
\includegraphics[width=9 cm,height=6cm]{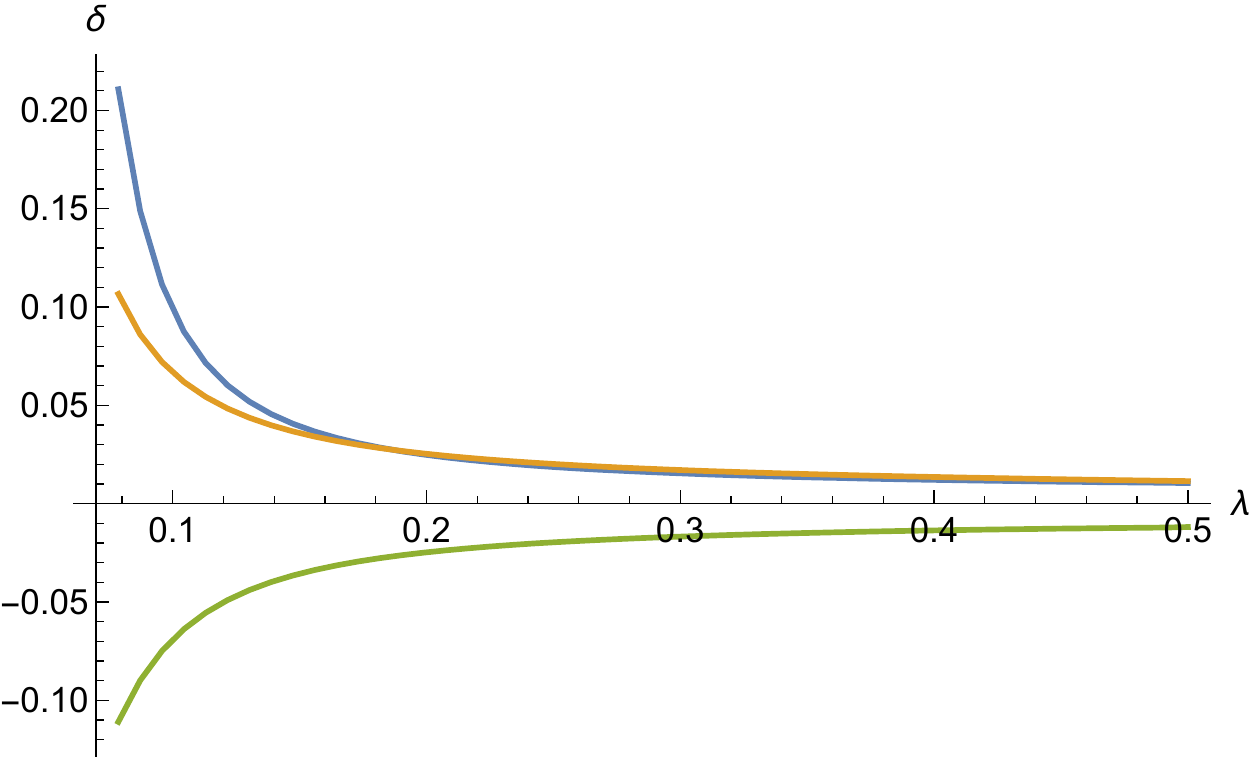}
\par\end{centering}
\caption{\label{newpl} 
Relative error $\delta = (\Delta E_{erg} - \Delta E_{exact})
/\Delta E_{exact}$ for the 
SA ERG (blue upper positive curve),
and for the simplified SA ERG with $\eta_k=0$
(orange lower positive curve),
and for the NSA ERG (green  negative curve),
vs. the coupling $\lambda$.
}
\end{figure}

Now we turn to the SA flow and show the output obtained in
the ERG case, i.e. from 
the flow equations (\ref{saERGV}), (\ref{saERGZ}),
that include non-linear effects in the 'time' derivatives,
due to the presence of $\eta_k$. Unfortunately, these non-linear terms
substantially affect the numerical convergence of the PDE, and we can 
trust the results of this scheme only for $\lambda\geq 0.08$.  
In order to get some insight, we report in Fig. \ref{newpl} 
the relative error 
on the determinations $\delta = (\Delta E_{erg} - \Delta E_{exact})
/\Delta E_{exact}$ as a function of the coupling $\lambda\geq 0.08$,
for the SA ERG (blue upper positive curve). 

The determination of 
$\delta$ for the NSA ERG (green negative curve) 
is also included for comparison.  
We observe that the relative error becomes
larger for  smaller values of the coupling in both cases, 
but  $\delta$  of the SA ERG is about twice the one of the NSA ERG
at $\lambda= 0.08$.
We also performed the analysis of the 
SA PT flow corresponding to the 'B-scheme' of \cite{bonannolippoldt}
but we do not report the corresponding curve, as
it essentially produces even worse results than 
those of the SA ERG scheme, shown by
the blue curve in Fig. \ref{newpl}.

Instead, the computation of the simplified SA ERG flow, 
where the nonlinear 
effects in Eqs. (\ref{saERGV}) and (\ref{saERGZ}) are discarded by
taking $\eta_k=0$, turns out to be much stabler than the full
SA (ERG and PT) flows. This is evident from the plot of the 
corresponding $\delta$ (red lower positive curve)
in Fig. \ref{newpl}, which is of the same 
size (although opposite in sign) of the NSA ERG relative error.  

\begin{figure}
\begin{centering}
\includegraphics[width=11 cm,height=8cm]{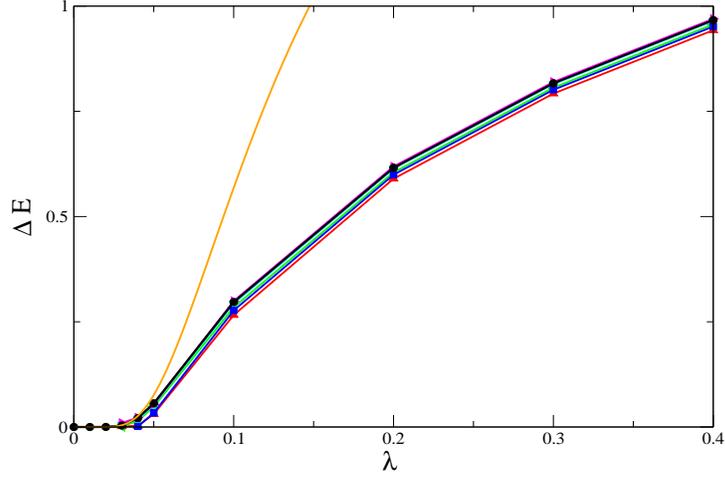}
\par\end{centering}
\caption{\label{figure4} Data reported in Table \ref{tab3}  
for $\Delta E$, as obtained by including the correction $Z_k$ 
in the simplified SA scheme,  plotted {\it vs.} $\lambda$. 
Again, for convenience, lines joining the points are included. Also the 
same coding as in Fig. \ref{figure1} is adopted, with the exception of 
$\Delta E_{WH}$ data, not included here as no $Z_k$ correction is 
available.}
\end{figure}

On the PT side, the flow that can consistently be compared with 
the SA ERG flow with $\eta_k=0$, corresponds to
the 'simplified type-C scheme' of \cite{bonannolippoldt},
reported in Eqs. (\ref{saPTV}), (\ref{saPTZ}).
In fact, we integrated these two SA simplified flows and, 
in addition to a small relative error, we also found in 
both cases convergence at smaller values of the coupling, 
so that  our analysis  could be pushed  down to $\lambda=0.03$. 
Therefore, below, we discuss
in more detail the output of these two sets of PDE, namely 
Eqs. (\ref{saERGV}), (\ref{saERGZ}) and (\ref{saPTV}), (\ref{saPTZ}),
as representative respectively of the ERG SA  and PT SA flow, and the 
relative data are reported in  Table \ref{tab3} and also collectively 
shown in Fig.\ref{figure4}.

\begin{table}[htb]
\begin{tabular}{|c | c | c | c | c | c | c | c |}\hline\hline
$M^2=1$ &&&&&&&\\
$\;\;\lambda\;\;$ & $\;\;\Delta E_{exact}\;\;$
&$\;\;\Delta E_{erg}\;\;$& $\;\;\Delta E_{pt1}\;\;$  & $\;\;\Delta E_{pt2}\;\;$&  $\;\;\Delta E_{pt3}\;\;$& $\;\;\Delta E_{pt4}\;\;$&   $\;\;\Delta E_{pt10}\;\;$\\
\hline\hline
 1   &1.9341 & 1.9249& 1.9197  &  1.9279 & 1.9311 & 1.9329 & 1.9362  \\
      &           & (1.009)& (1.010) &  (1.008) & (1.007)& (1.007) &  (1.006)  \\
\hline
0.02 &1.0540 &1.0539& 1.0538   &1.0540   &1.0541   & 1.0543   & 1.0542  \\
        &            &(1.003)& (1.0003) &(1.0003) &(1.0002)& (1.0002) & (1.0002) \\
\hline\hline\hline\hline 
$M^2=-1$&&&&&&&\\
$\;\;\lambda\;\;$ & $\;\;\Delta E_{exact}\;\;$
&$\;\;\Delta E_{erg}\;\;$& 
$\;\;\Delta E_{pt1}\;\;$  & $\;\;\Delta E_{pt2}\;\;$&  $\;\;\Delta E_{pt3}\;\;$& $\;\;\Delta E_{pt4}\;\;$&   $\;\;\Delta E_{pt10}\;\;$\\
\hline\hline
0.4   & 0.9667  & 0.9514 & 0.9427 & 0.9564  & 0.9617 & 0.9645 & 0.9699 \\
        &              & (1.040) & (1.048) & (1.035) & (1.031) & (1.029) &(1.024) \\
\hline
0.3   & 0.8166  &  0.8011 &0.7922 & 0.8062 & 0.8116  & 0.8144 &  0.8200 \\
        &              & (1.051) &  (1.062)& (1.045) & (1.039) & (1.036) & (1.031)\\
\hline
0.2  & 0.6159   & 0.5994  & 0.5899 &0.6049 &0.6105  & 0.6136 &  0.6192\\
       &               & (1.082)  &(1.101)& (1.071)& (1.061) & (1.056)& (1.047)\\
\hline
0.1  & 0.2969   & 0.2769  &0.2660 &0.2844 &0.2905 &0.2935 & 0.2994\\
       &               & (1.308) &(1.400) & (1.245)& (1.204)&(1.185) & (1.151)\\
\hline
0.05  & 0.0562 & 0.0334 & 0.0306&0.0483  &  0.0516 & 0.0527& 0.0539 \\
         &             & (5.872)  & (7.328)& (3.291)& (2.763) & (2.559) & (2.255)\\ 
\hline
0.04  & 0.0210 & 0.0013 &  0.0027 & 0.0141& 0.0167 &0.0178  & 0.0224 \\
         &             & (138.2) &  (80.51)& (8.649)& (6.252)& (5.397) & (3.162) \\ 
\hline
0.03  & 0.0036 & X& 0.0016 &0.0015 &  0.0036& 0.0059 &  0.0099 \\  
         &             &   &(90.71) & (14.67)& (11.27) & (6.51)  &  (3.85)  \\      
\hline\hline
\end{tabular}
\caption{Same as in Table \ref{tab2},
but with  $\Delta E$ derived from the
SA ERG flow in Eqs. (\ref{saERGV}), (\ref{saERGZ})
and SA PT flow in Eqs. (\ref{saPTV}), (\ref{saPTZ}).
}
\label{tab3}
\end{table}  
\normalsize

A comparison of Fig.\ref{figure4} and Fig.\ref{figure1} clearly 
indicates that 
the agreement of the SA data with $\Delta E_{exact}$  is improved with 
respect to the LPA data  in the region of small $\lambda$.
In addition, from Table \ref{tab3}  we find
$\Delta E_{pt1}<\Delta E_{erg}<\Delta E_{pt2}$, with the exception of 
the smallest values of the coupling $\lambda=0.04$ and $\lambda=0.03$,
which we shall comment on later. 
In both rows with $M^2=1$ in Table \ref{tab3}, we find
$\Delta E_{pt1}<\Delta E_{exact}<\Delta E_{pt10}$, 
but it must be remarked that the difference 
$\Delta E_{pt10} - \Delta E_{pt1}$, although smaller than $1\%$ 
in the two cases, is twice the 
corresponding difference for the NSA case. 

\begin{figure}
\begin{centering}
\includegraphics[width=13 cm,height=8cm]{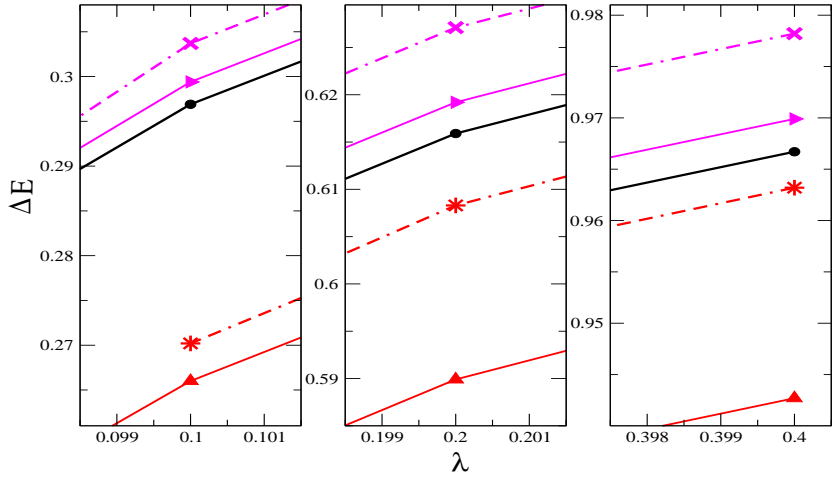}
\par\end{centering}
\caption{\label{figure5} Details of $\Delta E$, 
including the correction $Z_k$, at three values of the coupling  
$\lambda=0.1,\, 0.2 \, 0.4$. 
In addition to  $\Delta E_{exact}$ data from  Table \ref{tab0}
(black circles with a continuous black line),
we plot  $\Delta E_{pt1}$ and $\Delta E_{pt10}$
from  Table \ref{tab2} for the NSA case,
respectively with red stars  and pink crosses, both 
with  dot-dashed lines of the respective colour attached.
Moreover, $\Delta E_{pt1}$ and $\Delta E_{pt10}$
from  Table \ref{tab3} for the SA case,
are plotted with the same coding as in Figs.\ref{figure1} 
and \ref{figure4}, 
i.e. the former with red triangles  
pointing up  and  the latter with pink triangles pointing right. Both 
have continuous lines of the respective colour attached.}
\end{figure}

Turning to the $M^2=-1$ problem, a picture of the data obtained at 
$\lambda=0.4,\,0.2,\,0.1$ is given in the three panels of 
Fig.\ref{figure5} where, together with $\Delta E_{exact}$
(black circles),  the SA determination of $\Delta E_{pt1}$ 
(red triangles  pointing up) and of $\Delta E_{pt10}$ 
(pink triangles pointing right) 
and, in addition, also the NSA determination of $\Delta E_{pt1}$ 
(red stars) and of  $\Delta E_{pt10}$ (pink crosses) are displayed.
In all panels the point $\Delta E_{exact}$ is always between the 
$\Delta E_{pt10}$  and $\Delta E_{pt1}$ and the distance between these
two determinations is smaller for the NSA case at $\lambda=0.4$ but 
it becomes practically equal to the one for the SA case at $\lambda=0.1$
At smaller coupling $\lambda$, as already seen, the NSA flow fails to 
converge and only the SA determinations are available at
very low $\lambda$.

\begin{figure}
\begin{centering}
\includegraphics[width=11 cm,height=8cm]{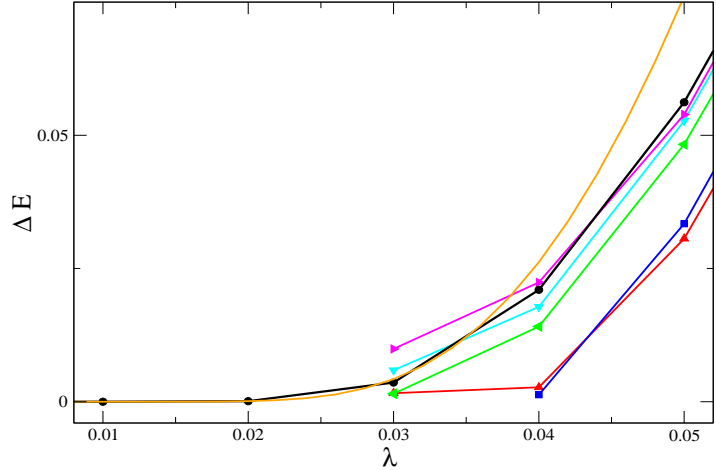}
\par\end{centering}
\caption{\label{figure6} 
Details at small values of  $\lambda$ of Fig.\ref{figure4}. $\Delta E$, 
including the correction $Z_k$, is plotted {\it vs.} $\lambda$. 
Coding is the same as in Figs.\ref{figure1}, \ref{figure4}
and \ref{figure5}.}
\end{figure}

Finally, the region of $\lambda\leq 0.05$ is displayed in
Fig.\ref{figure6}
where the SA determinations are plotted together with $\Delta E_{exact}$
and $\Delta E_{inst}$. 
Fig.\ref{figure6} clearly  shows the importance of the 
inclusion of $Z_{k=0}$ (which has now values of 
order 10 or in some cases  even 100) : points that in the LPA 
approximation are very far from  $\Delta E_{exact}$, 
now have substantially reduced their distance.

In particular, even the worst determination ($\Delta E_{pt1}$) at 
$\lambda= 0.05$ shows an agreement( about $35\%$) very similar 
to that of the instanton calculation.
At $\lambda= 0.04$, while $\Delta E_{pt1}$ and $\Delta E_{erg}$
become very small, calculations at larger $m$ (e.g. $m=10$)
are reasonably accurate, with error below $10\%$.

At $\lambda= 0.03$, $\Delta E_{inst}$ still differs from $\Delta 
E_{exact}$ 
of about $30\%$, while the PT determinations at $m=1$ and $m=2$ are off 
of $45\%$. The case with  $m=3$ (not reported in Fig.\ref{figure6} - 
see Table \ref{tab3}) practically reproduces $\Delta E_{exact}$, 
and $m=10$ yields too large values of the gap. Instead,
the ERG flow fails to converge at $\lambda= 0.03$, and 
this is probably due to an increasing $Z_k$ as $k\to 0$ 
(already at $\lambda= 0.04$ we find $Z_{k=0}=138$) that reduces 
the gap to such a  small value to be comparable with the 
numerical precision.

The same kind of problem is observed at $\lambda= 0.02$ not only for 
the ERG flow, but also for the PT flow with $m=1$ and $m=2$.
Yet, at $m=3,4,10$ we find convergence, respectively 
to  $3\;10^{-4}$,  $9\;10^{-4}$, $3.4 \;10^{-3}$ which, at least for 
$m=3$, is of the same order of magnitude of $\Delta E_{exact}$;
however we prefer not to report these values in Table \ref{tab3}
because, especially for the first two cases, they are comparable
with the numerical precision of our computation.

Therefore, we conclude that, when  including the correction 
$Z_{k=0}$, our computation  qualitatively reproduces 
the exponential trend induced by the tunnelling 
in the region $\lambda= 0.03-0.05$.

\section{Conclusions\label{conc}}

The computation of the energy gap in the quanto-mechanical 
double well potential by means of the derivative expansion
of the Functional Renormalization
Group flow equations performed in this paper, 
essentially aims at understanding to which extent this approach 
can quantitatively keep under control the simplest one-dimensional
non-perturbative effect, associated to the tunneling between 
the two vacua.  Therefore, we mainly focus on the 
comparison of  the  average estimates of various formulations 
of the flow equations with the accurate results coming from the 
Schroedinger equation resolution of the problem.

The improvement associated to the inclusion of higher terms 
of the derivative expansion is  evident even in the case of the
anharmonic oscillator with $M^2=1$, as shown in Tables 
\ref{tab1}, \ref{tab2}, \ref{tab3}. In this case,
the agreement of our estimates is always excellent (with a 
relative error that never exceeds $1\%$), but the improvement 
when going from the LPA in Table \ref{tab1} to the the NSA 
approximation in Table \ref{tab2} and simplified SA in Table
\ref{tab3}, is evident especially in the large coupling case 
$\lambda=1$.

Turning to the double well case with $M^2=-1$,
we can clearly distinguish different regimes 
associated to the  explored range $\lambda$.
Fig. \ref{figure1} shows that the estimates 
obtained in LPA for $\lambda \geq 0.2$ are globally accurate;
then , for
lower $\lambda$ the agreement diminishes and for $\lambda \geq 0.05$,
this approximation hardly reproduces the exponential 
approach to zero of $\Delta E$. As shown in Fig. \ref{figure3},
the WH approach has an inaccurate linear behavior, 
while for instance the PT potential with $m=10$ shows a 
qualitative agreement with the exact values.

Then, the inclusion of $Z_k$ in the simplified SA version of the flow
equations, clearly brings a strong improvement in the region 
$0.1 \leq \lambda \leq 0.03$ as seen in Figs. \ref{figure4} and 
\ref{figure6}. Even at $\lambda=0.04$, some estimates turn out to 
be more accurate than the instanton determination of $\Delta E$,
which is then no longer valid at $\lambda=0.03$ where 
$\Delta E_{inst}$ becomes the most precise estimate displayed.
However, it must be remarked that already at $\lambda = 0.04$,
and more evidently at smaller $\lambda$,
not all formulations of the flow equations show clear convergence
within our numerical precision. This drawback becomes too strong 
at $\lambda=0.02$ and we do not trust our findings at this 
value of the coupling.

To summarize, the inclusion of the first correction to the LPA 
represents a strong improvement in the range 
$0.1 \leq \lambda \leq 0.03$, indicating that the 
derivative expansion of the RG flow does 
actually reproduce the non-perturbative exponential 
decay of $\Delta E$. It is conceivable to expect 
that the next step in the derivative expansion, corresponding to
to inclusion of four derivatives in the original action,
could further enhance the convergence of the PDE for
$\lambda\leq 0.03$.

A final comment is dedicated to the different approaches adopted in 
our analysis. We found that, both NSA and SA (in either full or
simplified version) are equally accurate as long as the wave 
function renormalization $Z_k\sim 1$, as shown in Fig. \ref{newpl}.
But, as soon as $Z_k$ grows up to non perturbative values, the
NSA and the full SA schemes  become not trustable or do not converge,
the former because of the incomplete  cancellation of the term 
$Z_k p^2$ in the propagators, and the latter because of the 
presence of non-linear terms in the 'time' derivative of $Z_k$, 
included in $\eta_k$. In both cases the growth of $Z_k$ has
these undesired effects. Still, in the simplified SA approach,
where these problems are circumvented, PDE are integrated
at much lower values of $\lambda$ and,
at least in the PT flow with $m$ suitably optimized, 
providing precise  estimates of $\Delta E$.

\acknowledgements{This work has been partially supported by 
the INFN project FLAG. AB would like to thank Manuel Reichert for 
comments on the manuscript.}

\end{document}